\documentclass[letterpaper,12pt]{article}   
\usepackage{osajnl2} 
\usepackage[draft]{hyperref} 

\begin{document}

\title{A coherent optical link through the turbulent atmosphere}


\author{Khelifa Djerroud,$^{1}$ Ouali Acef,$^1$ Andr\'{e} Clairon,$^1$ Pierre Lemonde,$^1$ Catherine~N.~Man,$^2$ Etienne Samain,$^3$ and Peter Wolf$^{1,*}$}
\address{$^1$LNE-SYRTE, Observatoire de Paris, CNRS, UPMC, 61 av. de l'Observatoire, \\ 75014 Paris, France}
\address{$^2$ARTEMIS, Observatoire de la C\^{o}te d'Azur, Univ. Nice, CNRS, 06304 Nice Cedex, France}
\address{$^3$G\'{e}oAzur, Observatoire de la C\^{o}te d'Azur, Univ. Nice, CNRS, 06460 St Vallier de Thiey, France}
\address{$^*$Corresponding author: Peter.Wolf@obspm.fr}

\begin{abstract}
We describe the realization of a 5~km free space coherent optical
link through the turbulent atmosphere between a telescope and a
ground target. We present the phase noise of the link, limited
mainly by atmospheric turbulence and mechanical vibrations of the
telescope and the target. We discuss the implications of our
results for applications, with particular emphasis on optical
Doppler ranging to satellites and long distance frequency
transfer.
\end{abstract}

\ocis{010.1330, 120.5050, 060.2605.}

\maketitle 


Atomic clocks have been improving rapidly over the past years and
are now reaching uncertainties in relative frequency of 2 parts in
$10^{17}$ after less than $5\times10^4$ seconds of integration
time \cite{Rosenband}. Applications of such clocks in fundamental
physics, geodesy, navigation etc... require their comparison over
large distances without degrading their performance. In the longer
term one expects applications of such clocks on board terrestrial
and solar system satellites \cite{EGE,SAGAS}, which require a high
performance free space link from ground to space and/or between
spacecraft. At present, no existing long distance comparison
method reaches the required level of uncertainty \cite{Bauch}, and
even the improved microwave link of the ACES (Atomic Clock
Ensemble in Space) mission \cite{MWL} or the optical T2L2 (Time
Transfer by Laser Link) link \cite{T2L2} will require several days
of integration time to reach $10^{-17}$. Over short to medium
distances ($\approx 10^2$~km) optical fibre links have
demonstrated sufficient performance \cite{Giorgio}, and our aim is
to extend those methods to free space propagation and towards
ground to satellite and intercontinental (via a relay satellite)
frequency comparison of clocks. Other applications of coherent
free space links are optical satellite Doppler ranging and
broadband optical communications, the latter being the focus of
much attention in recent years with the promise of Gb/s data rates
over large distances. At such high rates the main limitation is
strong amplitude fluctuation (scintillation) due to atmospheric
turbulence, which has been investigated experimentally in a 142~km
coherent optical link between two Canary islands \cite{Perlot}. In
contrast, we focus on the low frequency ($< 1$~kHz) part of the
link phase noise spectrum, of relevance to clock comparisons and
optical Doppler ranging, which is to a large extent independent of
amplitude fluctuations. Recently a short baseline (10 m)
laboratory experiment \cite{Alatawi} and a roof-top experiment
over a 100~m distance have been reported \cite{Freestyle}. The
latter concludes that free-space coherent optical links may only
be suitable for short distance ($<$ 1~km) clock comparisons. We
arrive at the opposite conclusion showing not only that such links
display high performance over our 5~km distance, but also that
they hold great promise for satellite to ground links, with the
perspective of reaching the performance of the best optical clocks
in less than 1000~s integration time. In this letter we present a
brief description of the experiment and our main results, with
more detailed information relegated to a forthcoming publication.

Our experiment took place at the Observatoire de la C\^{o}te
d'Azur (OCA) lunar and satellite laser ranging facility located on
the plateau de Calern at an altitude of 1323 m, with data taken
during several days in June and July 2009. The set up consists of
a heterodyne Michelson interferometer with unequal arm lengths
(see fig. 1), using a 1064~nm Nd:YAG laser (Innolight/Prometheus,
$\approx 1$~kHz linewidth). The local arm is frequency shifted by
200~MHz using a Acousto-Optic Modulator (AOM) and recombined with
the distant arm on the photodiode. The resulting heterodyne beat
signal is used to phase lock a Voltage Controlled Oscillator (VCO)
with the locking bandwidth set to around 50~kHz. The frequency of
the VCO is counted using a zero dead-time counter with a 1~kHz
data rate. We also generate a quadrature signal (mixer M2) with
respect to the phase locked loop (mixer M1), which is proportional
to the beat signal amplitude but independent of its phase (to
first order) when the loop is closed. It thus allows monitoring of
the signal amplitude. The distant arm is fed into the Lunar Laser
Ranging 1.5 m aperture telescope at OCA. The beam diameter at the
exit of the telescope is about 380~mm. The distant arm is
reflected by a 5~cm corner cube mounted on an iron structure
$\approx$3.5~m off the ground on a mountain top $\approx$2.5~km
from the telescope. We feed $\approx$275~mW of optical power into
the telescope for a received average power on the photodiode of
$\approx$20~$\mu$W. The main sources of loss are the small size
(relative to the beam diameter) of the corner cube and the low
transmission of the telescope at 1064~nm (optimized for 532~nm).
As expected, we observe large amplitude fluctuations in the
heterodyne beat signal, caused by scintillation from atmospheric
turbulence.

\begin{figure}
\begin{center}
\includegraphics[width=8.3cm]{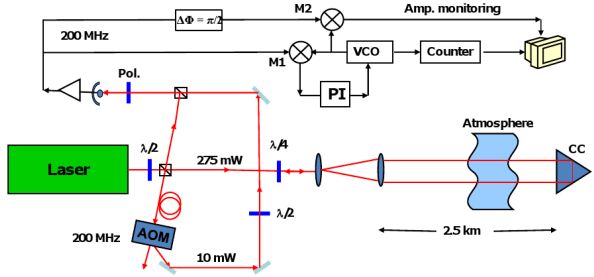}
\label{fig1} \caption{Principle of the experiment: AOM,
Acousto-Optic Modulator; Pol., Polarizer; PI,
Proportional-Integrator filter; VCO, Voltage Controlled
Oscillator.}
\end{center}
\end{figure}

In a ground-ground or ground-satellite free space coherent optical
link the main limitation in the low frequency part of the spectrum
is expected to arise from fluctuation of the refractive index of
the atmosphere due to turbulence. Models to estimate the phase
noise from turbulence can be found in \cite{Handbook} and
references therein. Using those models for our horizontal 2.5~km
return path we obtain a phase noise estimate about two orders of
magnitude worse than for a zenithal ground to satellite link from
the same location. Therefore, in order to obtain conditions
representative of a vertical link, we have taken data during the
calm periods of inversion of the temperature gradient (about 1~h
after sunrise) where the observed turbulence level was
significantly lower than during the rest of the day. This is also
confirmed by the fact that the observed noise during those calm
periods was close to the one reported in \cite{Linfield} obtained
by interferometric observation of light from stars crossing the
atmosphere vertically during average turbulence conditions. In
\cite{Linfield} turbulence is quantified by the phase noise
structure function defined as
$D_x(\tau)=\langle[x(t+\tau)-x(t)]^2\rangle$ where $x(t)$ is the
optical path length traveled by the signal received at time $t$.
The obtained structure functions for $0.1$~s $\leq\tau\leq 10$~s
can be described by a power law of the form $D_x(\tau)\simeq
C\tau^\beta$, with typical values in \cite{Linfield} of $D_x({\rm
1 s})\approx 20$~$\mu$m$^2$ and $\beta\approx 1.45$ and large
variations around those values. We typically obtain structure
functions following a similar power law with $D_x({\rm 1
s})\approx 25$~$\mu$m$^2$ and $\beta\approx 1.25$, which compare
well with the results of \cite{Linfield}. Both results show
significantly lower power law slopes than predicted from standard
Kolmogorov turbulence theory ($\beta=5/3$). We thus estimate that
the phase noise from turbulence on our horizontal link during calm
periods is representative (at least in order of magnitude) of the
phase noise expected in a one-way ground-satellite link in average
conditions.

Figure 2 shows the power spectral density (PSD) of fractional
frequency fluctuations for a 33~min data set taken on July 3
($\approx$1~h after sunrise) in clear sunny conditions. Similar
results are obtained for other days in similar conditions. The
lower curve represents our noise floor, obtained by placing a
mirror on the distant arm before injection into the telescope. It
is dominated at high Fourier frequency by the white phase noise of
our counter, and below 30~Hz by acoustic noise in the laboratory.
The upper curve is our measurement over the 5~km return distance
through the turbulent atmosphere. The initial data consists of
about $2 \times 10^6$ frequency measurements at 1~ms intervals,
with a standard deviation of 96~Hz. Optical cycle slips of the
phase-locked loop are easily identified, as at our 1~ms sampling
they correspond to 1~kHz steps. We removed 66 such points
$(\approx 3\times 10^{-5}$ of the data) all of which were obvious
cycle slips and correlated well with measured signal extinction
due to atmospheric scintillation. The PSD of our measurement shows
turbulence noise at frequencies below 10~Hz. For this part the PSD
is proportional to $f^{-0.3}$, below the expected value from
turbulence theory of $f^{-2/3}$ as already mentioned above. Above
10~Hz the PSD exhibits peaks due to mechanical resonances of the
telescope and the target with amplitudes around 1~$\mu$m, but we
also note an underlying increase of the PSD with frequency, due to
amplitude to phase noise conversion in our phase locked loop.

\begin{figure}
\begin{center}
\includegraphics[width=8.3cm]{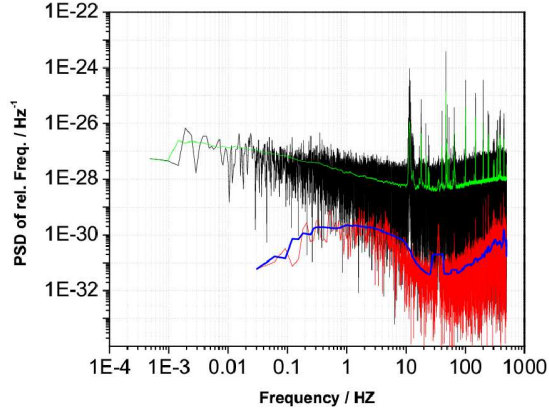}
\label{fig2} \caption{Fractional frequency power spectral density
of the free-space link (black) and system floor (red), with 500
points moving averages superposed in green and blue.}
\end{center}
\end{figure}

Figure 3 shows the Allan deviation of fractional frequency
($\sigma_y(\tau)$) of the same data set. At $\tau < 0.1$~s it is
dominated by the sum of the periodic effects due to mechanical
vibrations, and at larger $\tau$ its slope reflects the noise from
atmospheric turbulence ($\propto \tau^{-0.3}$), somewhat different
from the expected value from turbulence theory ($\propto
\tau^{-1/6}$), as already mentioned above and consistent with
measurements of \cite{Linfield}.


\begin{figure}
\begin{center}
\includegraphics[width=8.3cm]{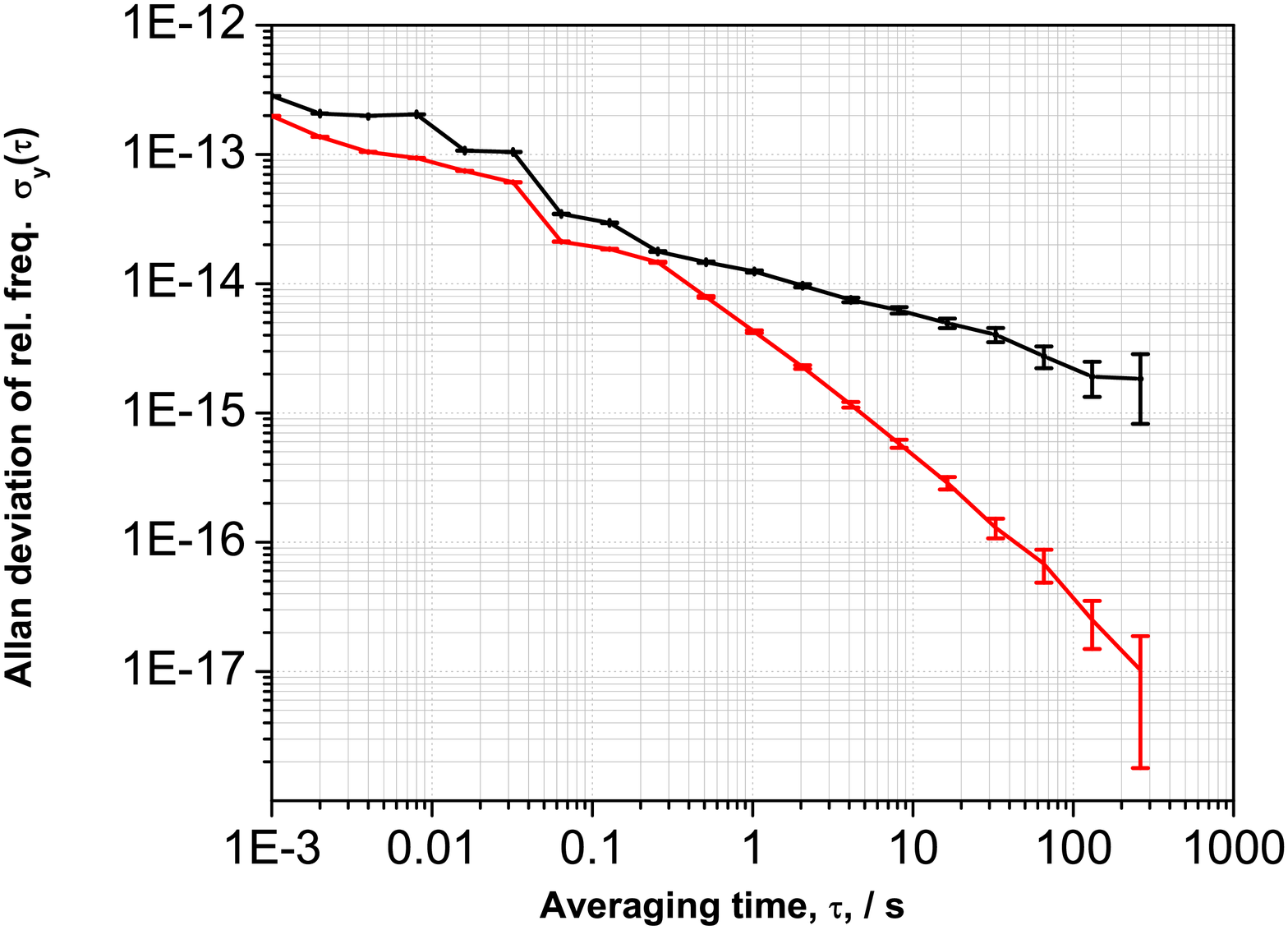}
\label{fig3} \caption{Fractional frequency stability {\bf
measured} on the ground-ground free space link (black), and {\bf
estimated} for a ground to geostationary satellite clock
comparison (red).}
\end{center}
\end{figure}

The Allan deviation of our data is about $1.3\times 10^{-14}$ at
$\tau=1$~s, and reaches $2 \times 10^{-15}$ after about 100~s
integration time, which is a remarkable stability for an
uncompensated link through the turbulent atmosphere. For instance,
it is similar to the uncompensated noise in the best $10^2$~km
fibre links \cite{Giorgio}. Given that, with appropriate two-way
compensation schemes, those links show the by far best performance
for frequency comparisons over short to medium distances, we
believe that if our results are representative of a
ground-satellite link, free space coherent optical links have
excellent potential for future ground to space and
intercontinental clock comparisons. This is discussed more
quantitatively below.

Let us assume a two-way free space optical link between a ground
station and a geostationary satellite, with heterodyne frequency
measurements of the received signal vs. the local signal, on board
($y_s(t_s)$), and on the ground ($y_g(t_g)$). The up and down
links will be affected by three main noise sources: satellite
motion, ground station motion, and atmospheric turbulence. We
further assume that the latter two are characterized by our
measured noise as shown above. The two way system is used in such
a way that the "up" signal is received at the satellite at the
time of emission of the "down" signal (to within a few ms), which
can easily be implemented during the data analysis when combining
the on board and ground measurements. The frequency difference of
the ground and on board optical clocks is given by $\Delta y =
(y_s(t_s) - y_g(t_g))/2$. In that difference the Doppler effect
from satellite motion cancels to a large extent (exactly if the
two signals are coincident at the satellite). However, the noise
from ground station motion and atmospheric turbulence partially
remains as the emission of the "up" signal is separated from the
reception of the "down" signal by the return travel time $\Delta t
\simeq 250$~ms. We can estimate that residual noise
$(y_{res}(t))$, by calculating $y_{res}(t) = (y(t) - y(t+\Delta
t))/2$ from our data, and investigate its statistics. The
resulting PSD is given by the one in fig. 2 multiplied by a
transfer function equal to $(1-{\rm cos}(2\pi f \Delta t))/2$,
resulting in a steep decrease at low frequency ($\leq 1$~Hz). The
corresponding Allan deviation of $y_{res}(t)$ is shown in fig. 3.
It stays a factor $\simeq \sqrt 2$ (due to the factor 2 in
$y_{res}$) below the measured noise at small $\tau$, but shows a
steep decrease starting at $\tau \approx 250$~ms, reaching
$1\times 10^{-17}$ after about 300~s integration time. Thus, free
space optical links hold great promise for future long distance
clock comparisons, with the potential of reaching the uncertainty
of the best present day optical clocks in less than 1000~s.
%

A limit to the above argument is imposed by signal outages due to
atmospheric scintillation and corresponding possible optical cycle
slips. These will be uncorrelated between the up and down links,
and will therefore fully affect $y_{res}(t)$. In our 33~min data
set we observed 66 such signal outages, which implies an average
loss of about 0.05 cycles per second. The resulting noise
corresponds to an Allan variance of $\sigma_y(\tau) \approx 2
\times 10^{-16}/\sqrt{\tau}$ still significantly better than the
best present optical clocks and just below the estimated stability
shown in Fig. 3. If these cycle slips are not random they cause a
frequency bias at worst equal to $\approx 2\times 10^{-16}$. This
does not affect the clock comparison as long as turbulence is
sufficiently low to allow identification and removal of cycle
slips (clearly the case in our data, see discussion above), and we
expect this to be the case at most astronomical observing sites.
Furthermore, signal outages can be mitigated by the use of
adaptive optics schemes.

For Doppler ranging the residual noise on the satellite Doppler
(and thus velocity) is given by $y_{res}(t) = (y(t) + y(t+\Delta
t))/2$. The Allan deviation of the corresponding residual distance
noise is $\sigma_x(\tau)= 28$~nm at $\tau=1$~ms and
$\sigma_x(\tau)= 1.4$~$\mu$m at $\tau=1$~s. Although this
corresponds to $>3$ orders of magnitude improvement on the present
measurement noise in satellite laser ranging, one should bear in
mind that, for lower orbits, the long term noise will be dominated
by changes of the tropospheric delay as the satellite passes
overhead (varying elevation), which can presently only be modeled
at the mm level at best.

In conclusion, we have demonstrated operation of a free space
coherent optical link through the turbulent atmosphere. Based on
our results we estimate that such links promise large improvements
in long distance clock comparisons and satellite ranging. We are
at present working towards extending our experiment to a
ground-space link using existing low orbit satellites equipped
with corner cube reflectors. The main challenges in doing so are
the large induced Doppler shift of our signal ($\pm 12$~GHz), the
low expected return power $< 1$~pW, and the longer signal travel
time which requires frequency stabilization of our laser.

{\bf Acknowledgements:} Valuable help by D. Albanese, S. Bize, M.
Lours, J. Paris, J. Pinto, G. Santarelli, J-M. Torre is gratefully
acknowledged.

\end{document}